\documentclass[aps,prd,onecolumn,superscriptaddress,nofootinbib,showpacs]{revtex4}

\usepackage{epsfig,amsfonts,amsthm}
\usepackage{amsmath,amssymb,color}
\usepackage{soul}
\usepackage{epstopdf}
\usepackage[utf8]{inputenc}
\usepackage[hidelinks]{hyperref}

\newcommand{\ba}{\begin{eqnarray}}
\newcommand{\ea}{\end{eqnarray}}
\newcommand{\be}{\begin{equation}}
\newcommand{\ee}{\end{equation}}

\begin{document}

\vspace*{-1.cm}
\begin{flushright}
{\small
DESY 18-129 \\
KA-TP-19-2018
}
\end{flushright}
\vspace*{1.5cm}

\title{CP in the Dark}

\author{Duarte Azevedo}
\email[E-mail: ]{dazevedo@alunos.fc.ul.pt}
\affiliation{Centro  de  F{\'i}sica  Te{\'o}rica  e  Computacional,  Universidade  de
Lisboa, 1649-003 Lisboa, Portugal}
\affiliation{LIP, Departamento de F\'{\i}sica, Universidade do Minho, 4710-057 Braga, Portugal}
\author{Pedro M. Ferreira}
\email[E-mail: ]{pmmferreira@fc.ul.pt}
\affiliation{Instituto Superior de Engenharia de Lisboa - ISEL, 1959-007 Lisboa,  Portugal}
\affiliation{Centro  de  F{\'i}sica  Te{\'o}rica  e  Computacional,  Universidade  de
Lisboa, 1649-003 Lisboa, Portugal}
\author{M.~Margarete Muhlleitner}
\email[E-mail: ]{margarete.muehlleitner@kit.edu}
\affiliation{Institute for Theoretical Physics, Karlsruhe Institute of Technology,
76131 Karlsruhe, Germany}
\author{Shruti Patel}
\email[E-mail: ]{shruti.patel@kit.edu}
\affiliation{Institute for Theoretical Physics, Karlsruhe Institute of Technology,
76131 Karlsruhe, Germany}
\affiliation{Institute for Nuclear Physics, Karlsruhe Institute of Technology,
76344 Karlsruhe, Germany}
\author{Rui Santos}
\email[E-mail: ]{rasantos@fc.ul.pt}
\affiliation{Instituto Superior de Engenharia de Lisboa - ISEL, 1959-007 Lisboa,  Portugal}
\affiliation{Centro  de  F{\'i}sica  Te{\'o}rica  e  Computacional,  Universidade  de
Lisboa, 1649-003 Lisboa, Portugal}
\affiliation{LIP, Departamento de F\'{\i}sica, Universidade do Minho, 4710-057 Braga, Portugal}
\author{Jonas Wittbrodt}
\email[E-mail: ]{jonas.wittbrodt@desy.de}
\affiliation{DESY, Notkestra{\ss}e 85, 22607 Hamburg, Germany}

\date{\today}

\begin{abstract}
We build a model containing two scalar doublets and a scalar singlet with a specific
discrete symmetry. After spontaneous symmetry breaking, the model has
Standard Model-like phenomenology, as well as a hidden scalar sector
which provides a viable dark matter candidate. We show that CP
violation in the scalar sector occurs exclusively
in the hidden sector, and consider possible experimental signatures of this
CP violation. In particular, we study contribution to anomalous gauge couplings
from the hidden scalars.
\end{abstract}

\maketitle

\section{Introduction}

The LHC discovery of a scalar with mass of $125$ GeV~\cite{Aad:2012tfa,Chatrchyan:2012xdj}
completed the Standard Model particle content. The fact that precision measurements of the
properties of this particle~\cite{Aad:2015zhl,Khachatryan:2016vau} indicate that it behaves
very much in a Standard Model (SM)-like manner is a further confirmation of the validity and
effectiveness of that model. Nonetheless, the SM leaves a lot to be explained, and many
extensions of the theory have been proposed to attempt to explain such diverse
phenomena as the existence of dark matter, the observed universal matter-antimatter asymmetry
and others. In particular, numerous SM extensions consist of enlarged scalar sectors, with
singlets, both real and complex, being added to the SM Higgs doublet \cite{McDonald:1993ex,Burgess:2000yq,O'Connell:2006wi,
BahatTreidel:2006kx,Barger:2007im,He:2007tt,Davoudiasl:2004be,Basso:2013nza,Fischer:2016rsh};
or doublets, the simplest example of which is the two-Higgs doublet model (2HDM)~\cite{Lee:1973iz,Branco:2011iw}.
Certain versions of singlet-doublet models provide dark matter candidates, as does the Inert version
of the 2HDM (IDM)~\cite{Deshpande:1977rw,Barbieri:2006dq,LopezHonorez:2006gr,Dolle:2009fn,
Honorez:2010re,Gustafsson:2012aj,Swiezewska:2012ej,Swiezewska:2012eh,Arhrib:2013ela,Klasen:2013btp,Abe:2014gua,
Krawczyk:2013jta,Goudelis:2013uca,Chakrabarty:2015yia,Ilnicka:2015jba}. Famously, the 2HDM was introduced in 1973 by
Lee to allow for the possibility of spontaneous
CP violation. But models with dark matter candidates {\em and} extra sources of CP violation
(other than the SM mechanism of CKM-matrix generated CP violation) are rare. Even rarer
are models for which a ``dark" sector exists, providing viable dark
matter candidates, and where the extra
CP violation originates exclusively in the ``dark" sector. To the best of our knowledge, the only
model with scalar CP violation in the dark sector is the recent work of
Refs.~\cite{Cordero-Cid:2016krd,Sokolowska:2017adz}, for which a three-doublet
model was considered. The main purpose of Refs.~\cite{Cordero-Cid:2016krd,Sokolowska:2017adz} was to describe the
dark matter properties of the model. In Ref.~\cite{TALK} an
argument was presented to prove that the model is actually
CP violating, adapting
the methods of Refs.~\cite{Grzadkowski:2016lpv,Belusca-Maito:2017iob}
for the complex 2HDM (C2HDM).

In the current paper we will propose a model, simpler than the one in~\cite{Cordero-Cid:2016krd}, but which boasts
the same interesting properties, to wit: (a) a SM-like Higgs boson, $h$, ``naturally" aligned due to the vacuum
of the model preserving a discrete symmetry; (b) a viable dark matter candidate, the stability of which is
guaranteed by the aforementioned vacuum and whose mass and couplings satisfy all existing dark matter
search constraints; and (c) extra sources of CP violation exist in the scalar sector of the model, but {\em only}
in the ``dark" sector. This {\em hidden} CP violation will mean that the SM-like scalar, $h$,
behaves almost exactly like the SM Higgs boson, and in particular (unless contributions from a high number of loops
are considered) $h$ has couplings to gauge bosons and fermions which are exactly those of a scalar. This is all
the more remarkable since the CP violation of the proposed model is {\em explicit}. The extra particle content
of the model, as advertised, is simpler than the model of~\cite{Cordero-Cid:2016krd}, consisting of two Higgs
doublets (both of
hypercharge $Y = 1$) and a real singlet ($Y = 0$). This is sometimes known as the Next-to-2HDM (N2HDM),
and was
the subject of a thorough study in~\cite{Muhlleitner:2016mzt}. The N2HDM version considered in this paper uses a different
discrete symmetry than the symmetries considered in~\cite{Muhlleitner:2016mzt}, designed, as will be shown, to produce
both dark matter
and dark CP violation. The paper is organised as follows: in section~\ref{sec:pot} we will introduce the model,
explaining in detail its construction and symmetries, as well as the details of spontaneous symmetry breaking
that occurs when one of the fields acquires a vacuum expectation value (vev). In section~\ref{sec:sca} we will
present the results of a parameter space scan of the model, where all existing constraints -- both theoretical
and experimental (from colliders and dark matter searches) -- are
taken into account; deviations from the SM behaviour
of $h$ in the diphoton channel, stemming from the existence of a charged scalar, will be discussed, as will
the contributions  of the model to dark matter observables; in section~\ref{sec:dcp} we will
discuss how CP violation arises in the dark sector, and how it might have a measurable impact in future colliders.
Finally, we conclude in section~\ref{sec:conc}.

\section{The scalar potential and possible vacua}
\label{sec:pot}

For our purposes, the N2HDM considered is very similar to that
discussed in Ref.~\cite{Muhlleitner:2016mzt},
in that the fermionic and gauge sectors are identical to the SM and the scalar sector is extended
 to include an extra doublet and also a singlet scalar field -- thus the model boasts two scalar doublets,
 $\Phi_1$ and $\Phi_2$, and a real singlet $\Phi_S$. As in the 2HDM, we will
require that the Lagrangian be invariant under a sign flip of some scalar fields, so that the number
of free parameters of the model is reduced and no tree-level flavour-changing neutral currents (FCNC)
occur~\cite{Glashow:1976nt,Paschos:1976ay}. The difference between the current
 work and that of~\cite{Muhlleitner:2016mzt} consists in the discrete symmetry applied to the Lagrangian --
 here, we will  consider a single $Z_2$ symmetry of the form
\be
\Phi_1 \,\rightarrow \,\Phi_1\;\;\;,\;\;\;
\Phi_2 \,\rightarrow \,-\Phi_2\;\;\;,\;\;\;
\Phi_S \,\rightarrow \, -\Phi_S\,.
\label{eq:z2z2}
\ee
With these requirements, the most general
scalar potential invariant under $SU(2)\times U(1)$ is given by
\ba
V &=& m_{11}^2 |\Phi_1|^2 \,+\, m_{22}^2 |\Phi_2|^2 \,+\, \frac{1}{2} m^2_S \Phi_S^2
\, +\, \left(A \Phi_1^\dagger\Phi_2 \Phi_S \,+\,h.c.\right)
\nonumber \\ & &
\,+\, \frac{1}{2} \lambda_1 |\Phi_1|^4
\,+\, \frac{1}{2} \lambda_2 |\Phi_2|^4
\,+\, \lambda_3 |\Phi_1|^2 |\Phi_2|^2
\,+\, \lambda_4 |\Phi_1^\dagger\Phi_2|^2\,+\,
\frac{1}{2} \lambda_5 \left[\left( \Phi_1^\dagger\Phi_2 \right)^2 + h.c. \right] \nonumber \\ & &
\,+\,\frac{1}{4} \lambda_6 \Phi_S^4 \,+\,
\frac{1}{2}\lambda_7 |\Phi_1|^2 \Phi_S^2 \,+\,
 \frac{1}{2} \lambda_8 |\Phi_2|^2 \Phi_S^2 \,,
\label{eq:pot}
\ea
where, with the exception of $A$, all parameters in the
potential are real.
As for the Yukawa sector, we consider all fermion fields {\em neutral} under this symmetry.
As such, only the doublet $\Phi_1$ couples to fermions, and the Yukawa Lagrangian is therefore
\be
-{\cal L}_Y\,=\, \lambda_t \bar{Q}_L \tilde{\Phi}_1 t_R\,+\,\lambda_b \bar{Q}_L \Phi_1 b_R
\,+\,\lambda_\tau \bar{L}_L \Phi_1 \tau_R\,+\,\dots
\label{eq:yuk}
\ee
where we have only written the terms corresponding to the third generation of fermions, with the
Yukawa terms for the remaining generations taking an analogous form. The
left-handed doublets for quarks and leptons are denoted by $Q_L$ and
$L_L$, respectively; $t_R$, $b_R$ and $\tau_R$ are the
right-handed top, bottom and $\tau$ fields; and $\tilde{\Phi}_1$ is the
charge conjugate of the doublet $\Phi_1$.

Notice that since the two doublets have the same quantum numbers and are not physical (only the mass
eigenstates of the model will be physical particles), the potential must be invariant under basis
changes on the doublets. This is a well-known property of 2HDMs, which the N2HDM inherits:
any unitary transformation of these fields, $\Phi_i^\prime = U_{ij}\Phi_j$ with a $2\times 2$
unitary matrix $U$, is an equally valid description of the theory. Though the theory is invariant
under such transformations, its parameters are not and undergo transformations dependent on $U$.
A few observations  are immediately in order:
\begin{itemize}
\item Since only $\Phi_1$ has Yukawa interactions it must have a vev to give mass to all charged fermions\footnote{And neutrinos as well, if one wishes to
consider Dirac mass terms for them.}.
In fact the Yukawa sector of this model is identical to the one of the
SM, and a CKM matrix arises there,
as in the SM.
\item The fact that all fermions
    couple to a single doublet, $\Phi_1$, automatically ensures that no scalar-mediated
tree-level FCNC occur, as in the 2HDM with a $Z_2$ symmetry~\cite{Glashow:1976nt,Paschos:1976ay}.
\item The $Z_2$ symmetry considered eliminates many possible terms in the potential,
but does {\em not} force all of the remaining ones to be real -- in particular, both the
quartic coupling $\lambda_5$ and the cubic one, $A$, can be {\em a priori} complex. However,
using the basis freedom to redefine doublets, we can absorb one of those complex phases into,
for instance, $\Phi_2$. We choose, without loss of generality, to render $\lambda_5$ real.
\end{itemize}

A complex phase on $A$ renders the model explicitly
CP violating. Considering, for instance,
the CP transformation of the scalar fields given by
\be
\Phi_1^{CP}(t,\vec{r})\,=\,\Phi_1^*(t,-\vec{r})\;\;\;,\;\;\;
\Phi_2^{CP}(t,\vec{r})\,=\,\Phi_2^*(t,-\vec{r})\;\;\;,\;\;\;
\Phi_S^{CP}(t,\vec{r})\,=\,\Phi_S(t,-\vec{r})\;\;\;,
\label{eq:cp}
\ee
we see that such a CP transformation, to be a symmetry of the potential, would require all of its parameters
to be real. Notice that the CP transformation of the singlet trivially does not involve
complex conjugation as $\Phi_S$ is real. In fact, this is a well-known CP property of singlet fields
\cite{Branco:1999fs}. One point of caution is in order: the complex phase of $A$ is not invariant under the
specific CP transformation of Eq.~\eqref{eq:cp}, but by itself that does not prove that the model
is explicitly CP violating. In fact, one could consider some form of generalized CP (GCP) transformation
involving, other than complex conjugation of the fields, also doublet redefinitions: $\Phi_i^{GCP}(t,\vec{r})
\,=\,X_{ij}\Phi_j^*(t,-\vec{r})$. The model can only be said to be
explicitly CP violating if  there does not exist
any CP transformation under which it is invariant. So, conceivably, though the model breaks the CP
symmetry defined by the transformation of Eq.~\eqref{eq:cp}, it might be invariant under some other one.
The point is moot, however: As we will see ahead, the vacuum of the model we will be considering is invariant under
the CP transformation of Eq.~\eqref{eq:cp} (and the
$Z_2$ symmetry of Eq.~\eqref{eq:z2z2}),
but the theory has CP violation. Thus the CP symmetry was broken to begin with, and hence the model is
explicitly CP violating.

Let us consider now the possibility of spontaneous symmetry breaking in which only the $\Phi_1$
doublet acquires a neutral non-zero vev: $\langle \Phi_1 \rangle = (0,v/\sqrt{2})^T$. Given the structure of
the potential in Eq.~\eqref{eq:pot}, the minimisation conditions imply that this is a possible solution,
with all scalar components of the doublets (except the real, neutral
one of $\Phi_1$) and the singlet equal to zero,
provided that the following condition is obeyed:
\be
m^2_{11}\,+\,\frac{1}{2}\lambda_1\,v^2\,=\,0\,.
\label{eq:min}
\ee
Since all fermion and
gauge boson masses are therefore generated by $\Phi_1$, it is mandatory that $v = 246$ GeV as usual.
At this vacuum, then, it is convenient to rewrite the doublets in
terms of their component fields as
\be
\Phi_1\,=\,\left( \begin{array}{c} G^+ \\ \frac{1}{\sqrt{2}}  (v + h \,+\, \mbox{i} G^0)\end{array}\right)
\;\;\; , \;\;\;
\Phi_2\,=\, \left( \begin{array}{c} H^+ \\ \frac{1}{\sqrt{2}}(\rho \,+\, \mbox{i} \eta)\end{array}\right)
\, ,
\label{eq:doub}
\ee
where $h$ is the SM-like Higgs boson, with interaction vertices with fermions and gauge bosons identical to those
expected in the SM (the diphoton decay of $h$, however, will differ from its SM counterpart). The mass of the $h$
field is found to be given by
\be
m^2_h \,=\,\lambda_1\,v^2\,,
\label{eq:mh}
\ee
and since $m_h = 125$ GeV, this fixes the value of one of the quartic couplings, $\lambda_1 \simeq 0.258$. The
neutral and charged Goldstone
bosons $G^0$ and $G^+$, respectively, are found to be massless as expected, and the
squared mass of the charged scalar $H^+$ is given by
\be
m^2_{H^+}\,=\,m^2_{22}\,+\,\frac{\lambda_3}{2}\,v^2 \;.
\ee
Finally, the two neutral components of the doublet $\Phi_2$, $\rho$ and $\eta$, mix with
the singlet component $\Phi_s \equiv s$ yielding a $3\times
3$ mass matrix,
\be
\left[M^2_N\right]\,=\,\left( \begin{array}{ccc}
m^2_{22}\,+\,\frac{1}{2}\bar{\lambda}_{345}\,v^2 & 0 & -\mbox{Im}(A)\,v \\
0 & m^2_{22}\,+\,\frac{1}{2}\lambda_{345}\,v^2 & \mbox{Re}(A)\,v \\
-\mbox{Im}(A)\,v & \mbox{Re}(A)\,v & m^2_S\,+\,\frac{1}{2}\lambda_7\,v^2
\end{array}\right)\,,
\label{eq:mn}
\ee
with $\bar{\lambda}_{345} = \lambda_3 + \lambda_4 - \lambda_5$ and
$\lambda_{345} = \lambda_3 + \lambda_4 + \lambda_5$.
There are therefore three neutral scalars other than $h$, which we call $h_1$, $h_2$ and $h_3$,
in growing order of their masses.
This mass matrix can then be diagonalized by an orthogonal rotation matrix $R$, such that
\be
R\,M^2_N\,R^T\;=\; \mbox{diag}\left(m^2_{h_1}\,,\,m^2_{h_2}\,,\,m^2_{h_3}\right)
\ee
and the connection between the original fields and the mass eigenstates is given by
\be
\left(\begin{array}{c} h_1 \\ h_2 \\ h_3 \end{array} \right)\;=\; R\,
\left(\begin{array}{c} \rho \\ \eta \\  s \end{array} \right)\,.
\ee
The rotation matrix $R$ can be parametrized in terms of three angles, $\alpha_1$, $\alpha_2$
and $\alpha_3$, such that
\be
R\,=\,\left( \begin{array}{ccc}
c_{\alpha_1} c_{\alpha_2} & s_{\alpha_1} c_{\alpha_2} & s_{\alpha_2}\\
-(c_{\alpha_1} s_{\alpha_2} s_{\alpha_3} + s_{\alpha_1} c_{\alpha_3})
& c_{\alpha_1} c_{\alpha_3} - s_{\alpha_1} s_{\alpha_2} s_{\alpha_3}
& c_{\alpha_2} s_{\alpha_3} \\
- c_{\alpha_1} s_{\alpha_2} c_{\alpha_3} + s_{\alpha_1} s_{\alpha_3} &
-(c_{\alpha_1} s_{\alpha_3} + s_{\alpha_1} s_{\alpha_2} c_{\alpha_3})
& c_{\alpha_2}  c_{\alpha_3}
\end{array} \right)\,,
\label{eq:matR}
\ee
where for convenience we use the notation $c_i = \cos \alpha_i$, $s_j = \sin \alpha_j$.
Without loss of generality, we may take the angles $\alpha_i$ in the interval $[-\pi/2\,,\,\pi/2]$.

In the following we discuss several phenomenological properties of this model.
The vacuum preserves the  $Z_2$ symmetry. As a result, the physical eigenstates
emerging from $\Phi_2$ and $\Phi_S$, {\em i.e.} the charged scalar $H^\pm$ and the neutral ones
$h_1$, $h_2$ and $h_3$, carry a quantum number -- a ``dark charge" equal to $-1$ --
which is preserved in all interactions, to all orders of perturbation theory.
In the following we refer to these four
eigenstates as ``dark particles''.
 On the other hand, the SM-like particles ($h$, the gauge bosons and all fermions) have
``dark charge" equal to 1.
The preservation of this quantum number means that dark particles must always be produced
in pairs while in their decays they must always produce at least one dark particle.
  Therefore,
the lightest of these dark particles -- which we will choose in our parameter scans to be
the lightest neutral state, $h_1$ -- is {\em stable}. Thus, the model provides one dark
matter candidate.

The model indeed shares many features with the Inert version of the 2HDM, wherein all particles
from the ``dark doublet" are charged under a discrete symmetry, the lightest of which is stable.
The main difference with the current model is the mixing that occurs between the two neutral components
of the doublet and the singlet due to the cubic coupling $A$, which can be appreciated
from the mass matrix of Eq.~\eqref{eq:mn}. In what concerns the charged scalar, though, most of the
phenomenology of this model is equal to the Inert 2HDM.

\section{Parameter scan, the diphoton signal and dark matter observables}
\label{sec:sca}

With the model specified, we can set about exploring its available parameter space, taking
into account all of the existing theoretical and experimental constraints. We performed a
vast scan over the parameter space of the model (100.000 different combinations of the parameters
of the potential of Eq.~\eqref{eq:pot}), requiring that:
\begin{itemize}
\item The correct electroweak symmetry breaking occurs, and the correct value for
the mass of the observed Higgs boson is obtained; as already explained, this is achieved
by requiring that $v = 246$ GeV in Eq.~\eqref{eq:doub} while at the same time the
parameters of the model are such that Eqs.~\eqref{eq:min} and~\eqref{eq:mh} are satisfied.
\item By construction, all tree-level interactions and vertices of the Higgs particle $h$
are identical to those of the SM Higgs boson.
As a consequence, all LHC production cross sections for $h$ are
identical to the values expected in the SM. Additionally, all decay
widths of $h$, apart from the diphoton case to be treated explicitly
below, are identical to their SM values up to electroweak corrections. This
statement holds as we require the $h_1$ mass to be larger than roughly
70 GeV, to eliminate the possibility of the decay $h\rightarrow h_1
h_1$ (when this decay channel is open it tends to affect the branching
ratios of $h$, making it difficult to have $h$ be SM-like).
\item The quartic couplings of the potential cannot be arbitrary. In particular, they must be such that
the theoretical requirements of boundedness from below (BFB) -- that the potential always tends to $+\infty$ along
any direction where the scalar fields are taking arbitrarily large values -- and perturbative unitarity --
that the model remains both perturbative and unitary,
in all $2\rightarrow 2$ scalar scattering processes  -- are satisfied. The model considered in the current paper
differs from the N2HDM discussed in Ref.~\cite{Muhlleitner:2016mzt} only via the cubic
coefficient $A$, so the
tree-level BFB and perturbative unitarity constraints described there (in sections 3.1 and 3.2) are exactly the ones
we should use here.
\item The constraints on the scalar sector arising from the Peskin-Takeuchi electroweak precision parameters
$S$, $T$ and $U$~\cite{Peskin:1990zt,Peskin:1991sw,Maksymyk:1993zm}
are required to be satisfied in the model. Not much of the parameter space is eliminated due to this constraint,
but it is still considered in full.
\item Since the charged scalar $H^\pm$ does not couple to fermions, all $B$-physics bounds usually
constraining its interactions are automatically satisfied. The direct LEP bound of $m_{H^\pm} > 90$ GeV assumed
a 100 \% branching ratio of $H^\pm$ to fermions, so that this constraint also needs not be considered
  here.
\item The dark matter observables were calculated using
\texttt{MicrOMEGAs}~\cite{Belanger:2006is,Belanger:2013oya} and compared to the results from Planck~\cite{Aghanim:2018eyx} and XENON1T~\cite{Aprile:2018dbl}.
\item Since all scalars apart from $h$ do not couple to fermions, no electric dipole moment constraints
need be considered, this despite the fact that CP violation occurs in the model.
\end{itemize}

With these restrictions, the scan over the parameters of the model was such that:
\begin{itemize}
\item The masses of the neutral dark scalars $h_1$ and $h_2$ and the charged one, $H^\pm$, were chosen
to vary between 70 and 1000 GeV. The last neutral mass, that of $h_3$, is obtained from the
remaining parameters of the model as explained in~\cite{Muhlleitner:2016mzt}.
\item The mixing angles of the neutral mass matrix, Eq.~\eqref{eq:matR}, were chosen at random
in the interval $-\pi/2$ and $\pi/2$.
\item The quartic couplings $\lambda_2$ and $\lambda_6$ are constrained, from BFB constraints, to
be positive, and were chosen at random in the intervals $[0\,,\,9]$ and $[0\,,\,17]$, respectively.
$\lambda_8$ is chosen in the interval $[-26\,,\,26]$.
\item The quadratic parameters $m^2_{22}$ and $m^2_S$ were taken between 0 and $10^6$ GeV$^2$.
\end{itemize}
All other parameters of the model can be obtained from these using the
expressions for the masses of the scalars and the definition of the matrix $R$.
The scan ranges for the quartic couplings are chosen larger than the maximally allowed ranges after imposing unitarity and BFB. Therefore, all of the possible values for these parameters are sampled.
We have used the implementation of the model, and all of its theoretical
constraints, in {\tt ScannerS}~\cite{Coimbra:2013qq}. {\tt N2HDECAY}~\cite{Engeln:2018mbg}, a code based on
{\tt HDECAY}~\cite{Djouadi:1997yw,Djouadi:2018xqq}, was
used for the calculation of scalar branching ratios and total widths, as in~\cite{Muhlleitner:2016mzt}.

As we already explained, the tree-level interactions of $h$ are
identical to the ones of a SM Higgs boson of identical
mass. The presence of the charged scalar $H^\pm$, however, changes the diphoton decay width of $h$, since a
new loop, along with those of the $W$ gauge boson and charged fermions, contributes to that width. This is identical
to what occurs in the Inert model, and we may use the formulae
of, for instance, Ref.~\cite{Swiezewska:2012ej}.
Thus we find that the diphoton decay amplitude in our model is given by
\be
\Gamma(h\rightarrow \gamma\gamma)\,=\,\frac{G_F \alpha^2 m^3_h}{128\sqrt{2} \pi^3}\,
\left|\sum_f N_{c,f} Q^2_f A_{1/2}\left(\frac{4 m^2_f}{m^2_h}\right) \,+\,A_1
\left(\frac{4 m^2_W}{m^2_h}\right)
\,+\,\frac{\lambda_3 v^2}{2 m^2_{H^\pm}}\,A_0
\left(\frac{4 m^2_{H^\pm}}{m^2_h}\right)
\right|^2\,,
\label{eq:ampli}
\ee
where the sum runs over all fermions (of electric charge $Q_f$ and number of colour degrees
of freedom $N_{c,f}$) and $A_0$, $A_{1/2}$ and $A_1$ are the well-known form factors
for spin 0, 1/2 and 1 particles (see for instance Refs.~\cite{Spira:1997dg,Spira:2016ztx}).
The charged Higgs contribution to the diphoton amplitude in
Eq.~\eqref{eq:ampli} changes this decay width,
and therefore the total decay width, hence all branching ratios, of $h$ with respect to
the SM expectation. However, the diphoton decay width being so small compared to the main decay channels
for $h$ ($b\bar{b}$, $ZZ$ and $WW$), the overall changes of the total $h$ width are minimal. In fact,
numerical checks for our allowed parameter points have shown that the branching ratios of $h$ to $b\bar{b}$, $\tau\bar{\tau}$,
$ZZ$ and $WW$ change by less than 0.05\% compared to the corresponding SM quantities -- therefore, all current
LHC constraints for the observed signal rates of $h$ in those channels
are satisfied at the 1$\sigma$ level.

\begin{figure}[t]
\centering
\includegraphics[height=8cm,angle=0]{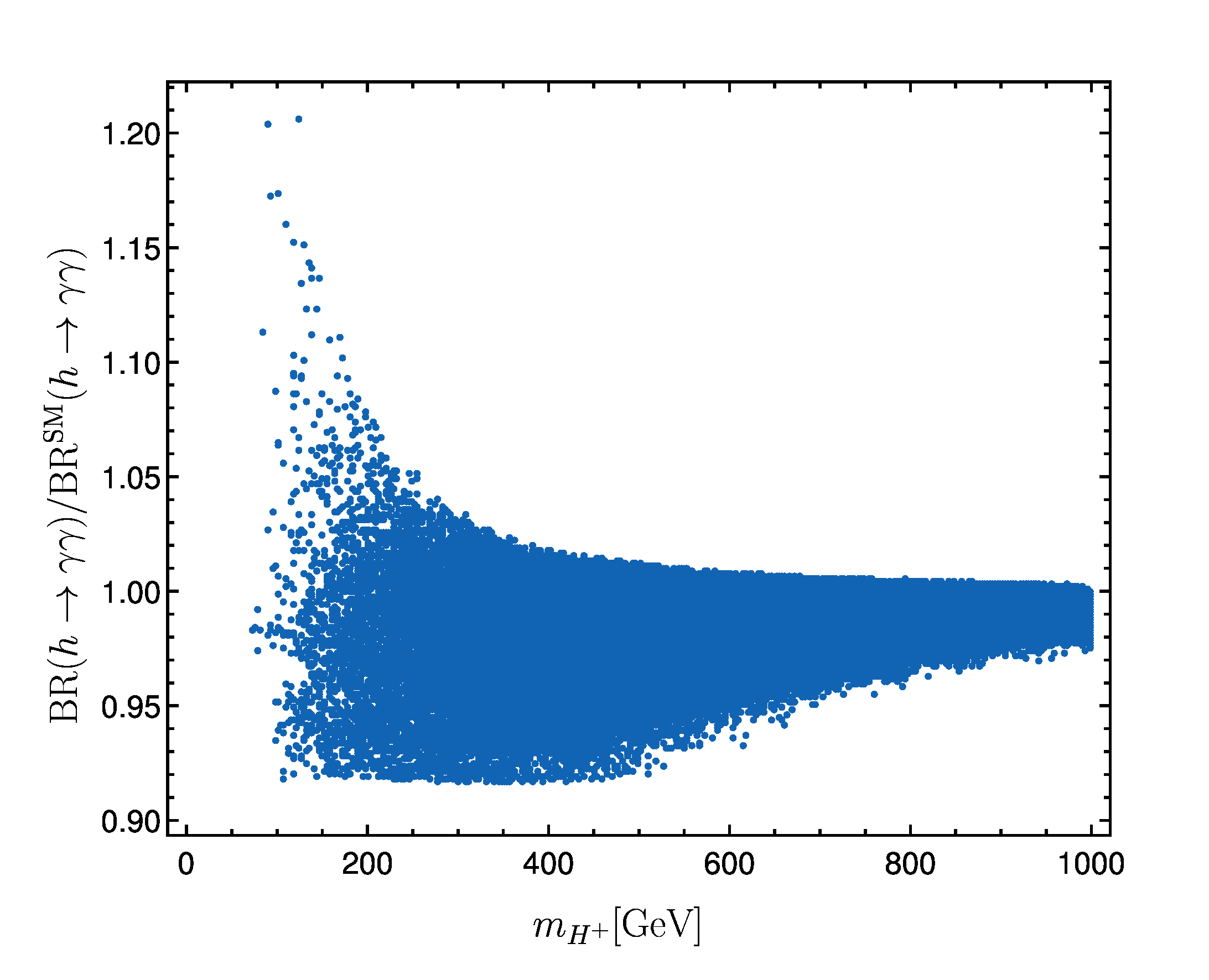}
\caption{Ratio of the branching ratio of $h$ into two photons to the SM value {\em versus}
the value of the charged scalar mass for all the
  allowed points in the model.
}
\label{fig:diph}
\end{figure}

As for the branching ratio
into two photons, it can and does change by larger amounts, as
can be appreciated from Fig.~\ref{fig:diph}.
In that figure we plot the ratio of the branching ratio of $h$ into two photons to its SM value as a function of the charged Higgs mass.
Comparing these results to the recent measurements of the $h\rightarrow \gamma\gamma$ signal
rates\footnote{Notice that since $h$ in this model has exactly the same
production cross sections as the SM Higgs boson, the ratio of branching
ratios presented in Fig.~\ref{fig:diph} corresponds
exactly to the measured signal rate, which involves the ratio of the product of production
cross sections and decay branching ratios, between observed and SM
theoretical values.} $\mu_{\gamma\gamma}$ from
Ref.~\cite{Sirunyan:2018ouh}, we see that our model can accommodate
values well within the 2$\sigma$ interval.
The lower bound visible in Fig.~\ref{fig:diph} emerges from the present experimental lower limit from~\cite{Sirunyan:2018ouh} at $2\sigma$.
The experimental upper limit, however, is larger than the maximum value of $\sim 1.2$ possible in our model.
The latter results from the combination of BFB and unitarity bounds which constrain the allowed values
of the coupling $\lambda_3$.
The lowest allowed value for $\lambda_3$, which governs the coupling of $hH^+H^-$, is about $-1.03$, and
its maximum one roughly $8.89$.
Since the value of $\mu_{\gamma\gamma}$ grows for negative $\lambda_3$, the lower bound on $\lambda_3$
 induces an upper bound of $\mu_{\gamma\gamma}\lesssim 1.2$.

Thus we see that the model under study in this paper is perfectly capable of reproducing
the current LHC data on the Higgs boson. Specific predictions for the diphoton signal rate are also
possible in this model -- values of $\mu_{\gamma\gamma}$ larger or smaller
than unity are easily accommodated, though they are constrained to the interval
$0.917 \lesssim \mu_{\gamma\gamma} \lesssim 1.206$.
As the parameter scan was made taking into account all data from dark matter searches, we are comfortable
that all phenomenology in that sector is satisfied by the dark
particles.

Let us now study how the model behaves in terms of dark matter variables.
\begin{figure}[h!]
\begin{tabular}{ccc}
\includegraphics[height=7cm,angle=0]{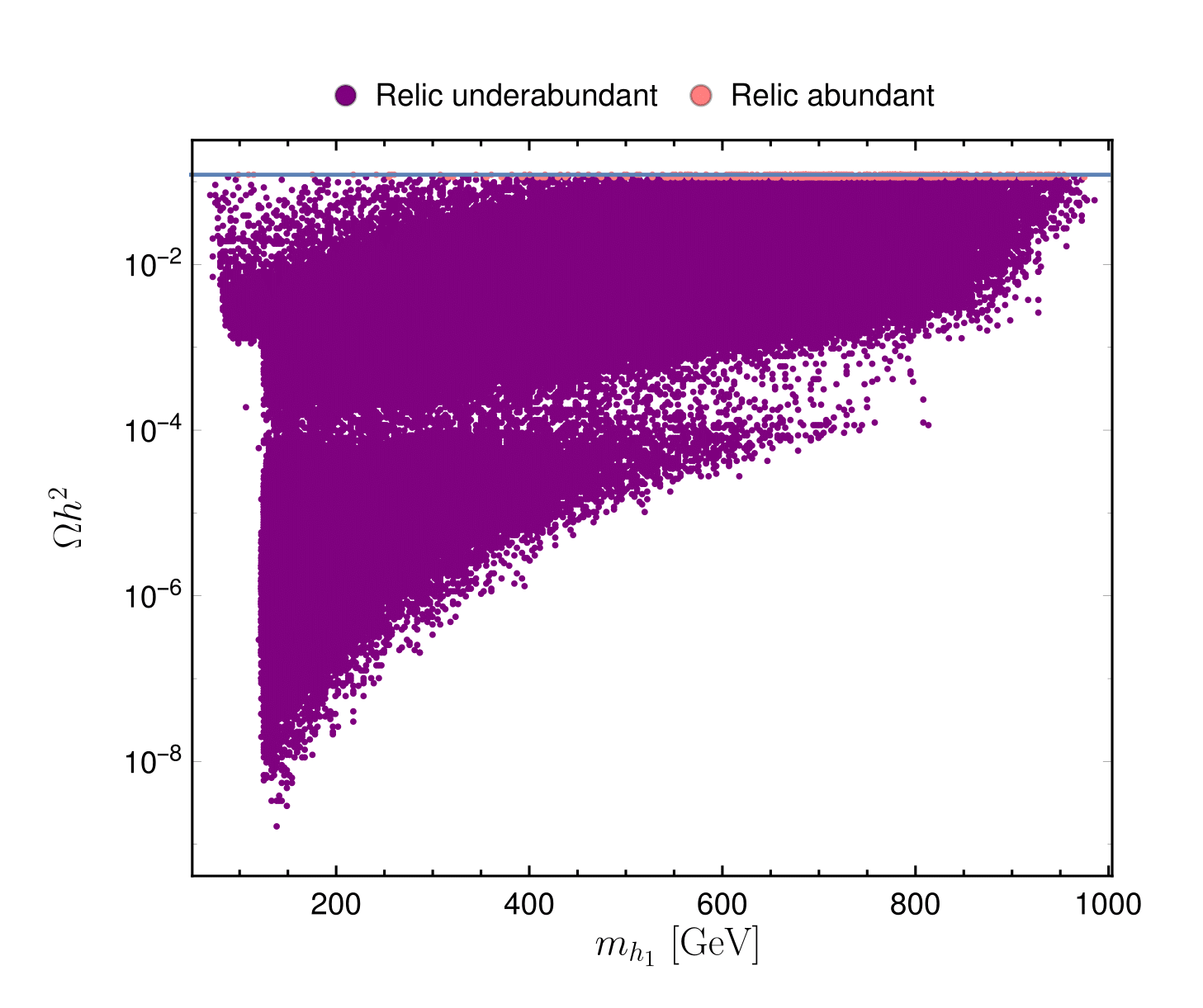}&
\includegraphics[height=7cm,angle=0]{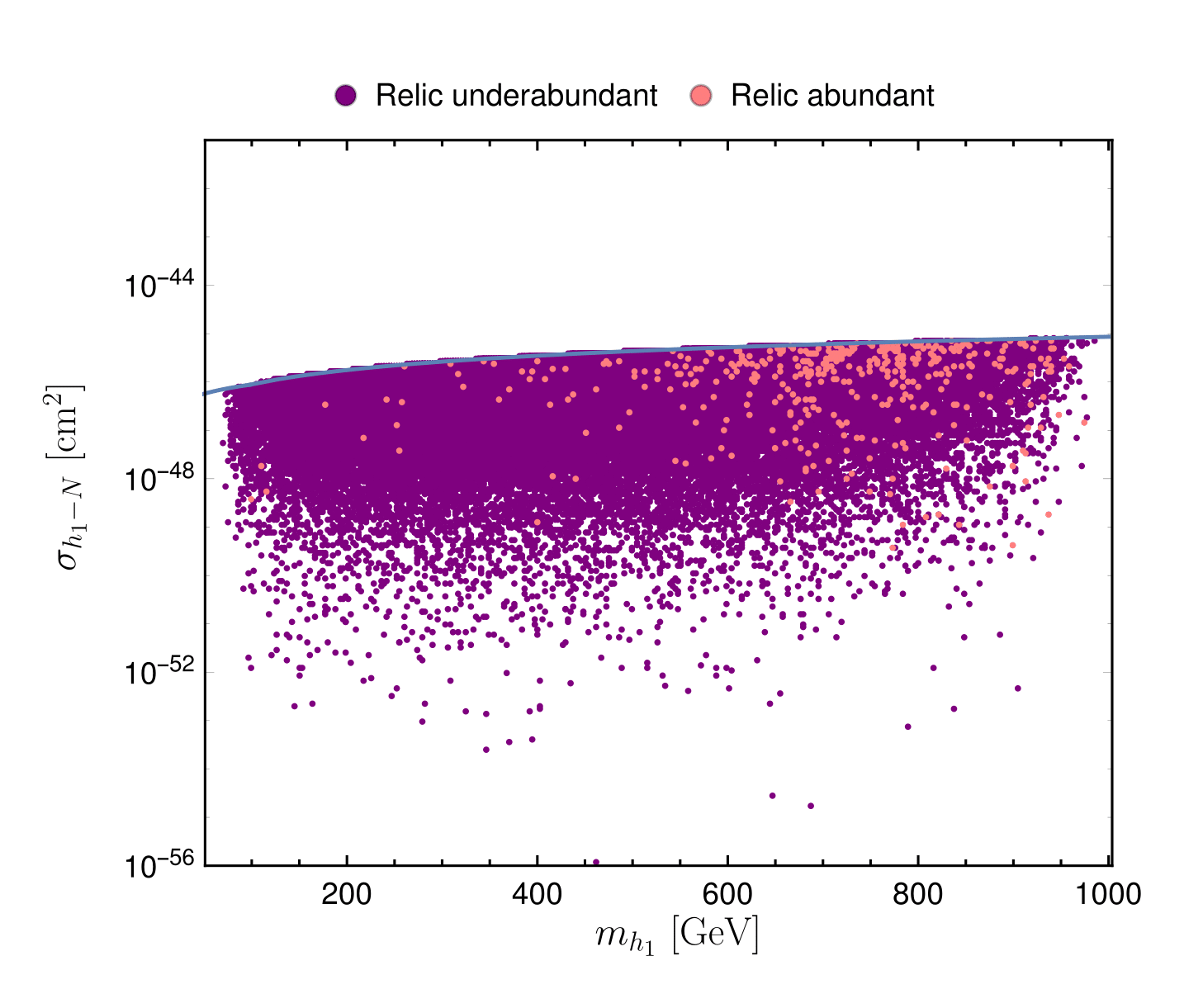}
\end{tabular}
\caption{Points that survive all experimental and theoretical constraints. Left: relic density abundance
versus dark matter mass where the grey line represents the measured
DM relic abundance; points either saturate the relic abundance
constraints within +1$\sigma$ and -$5 \sigma$ around the central value (pink points)
or are below the measured central value (violet points). Right: spin-independent nucleon
dark matter scattering cross section as a function of the dark matter
mass where the grey line represents
the latest XENON1T~\cite{Aprile:2017iyp,Aprile:2018dbl} results; colour code is the same and pink points are superimposed on violet points.
}
\label{fig:dm}
\end{figure}
Several experimental results put constraints on the mass of the dark matter (DM) candidate, and on its couplings to SM
particles. The most stringent bound comes from the measurement
of the cosmological DM relic abundance from the latest results from the Planck Collaboration~\cite{Aghanim:2018eyx},
 $({\Omega}h^2)^{\rm obs}_{\rm DM} = 0.120 \pm 0.001$.
The DM relic abundance for our model was calculated with \texttt{MicrOMEGAs}~\cite{Belanger:2013oya}. In our scan we accepted
all points that do not exceed the value measured by Planck by more than $1\sigma$. This way, we consider not only the
points that are in agreement with the DM relic abundance experimental values but also
the points that are underabundant and would need further dark matter candidates to saturate the measured experimental
value.

Another important constraint comes from direct detection experiments , in which the elastic scattering of DM off nuclear
targets induces nucleon recoils that are being measured by several experimental groups. Using the expression for
the spin-independent DM-nucleon cross section given by  \texttt{MicrOMEGAs},
we impose the most restrictive upper bound on this cross section, which is the one from
XENON1T~\cite{Aprile:2017iyp,Aprile:2018dbl}.

In the left panel of Fig.~\ref{fig:dm} we use the parameter scan previously described to compute dark matter observables.
We show the points that passed all experimental and theoretical constraints
in the relic abundance versus dark matter mass plane. We present in pink the points that saturate the relic abundance,
that is the points that are in the interval between $+1\sigma$ and $-5 \sigma$ around the central value, and
in violet the points for which the relic abundance is below the measured value.
It is clear that there are points in the chosen dark
matter mass range that saturate the relic density. In the right panel we present the spin-independent nucleon dark
matter scattering cross section
as a function of the dark matter mass. The upper bound (the grey line) represents the latest
XENON1T~\cite{Aprile:2017iyp,Aprile:2018dbl} results.
The pink points in the right plot show that even if the direct bound improves by a few orders of magnitude there
will still be points for the entire mass range where the relic density is saturated.

Thus we see that the model under study in this paper can fit, without need for fine tuning, the existing
dark matter constraints. Next we will study the rise of CP violation in the dark sector.

\section{CP violation in the dark sector}
\label{sec:dcp}

As we explained in section~\ref{sec:pot}, the model explicitly breaks the CP symmetry defined
in Eq.~\eqref{eq:cp}. Notice that the vacuum of the model which we are studying -- wherein
only $\Phi_1$ acquires a vev -- preserves that symmetry. Therefore, if there is CP violation (CPV)
in the interactions of the physical particles of the model, it did not arise from any spontaneous
CPV, but rather the explicit CP breaking mentioned above\footnote{Again, because this is a
subtlety of CP symmetries, let us repeat the argument: The fact that the model explicitly violates one
CP symmetry -- that defined in Eq.~\eqref{eq:cp} -- does not necessarily mean there is CPV, since
the Lagrangian could be invariant under a
different CP symmetry. If, however,  we prove that there is CPV
after spontaneous symmetry breaking with a vacuum that preserves the CP symmetry of Eq.~\eqref{eq:cp}, then
that CPV is explicit.}.

There are several eventual experimental observables where one could conceivably observe CPV. For instance,
a trivial calculation shows that all vertices of the form $Z h_i h_j$, with $i\neq j$, are possible. These
vertices arise from the kinetic terms for $\Phi_2$ where from
Eq.~\eqref{eq:doub} we obtain, in terms of
the neutral components of the second doublet,
\be
|D_\mu \Phi_2|^2\,=\, \dots \,+\,\frac{g}{\cos\theta_W}\,Z_\mu\left(\eta_2\partial^\mu \rho_2 \,-\,
\rho_2\partial^\mu \eta_2\right)\; ,
\ee
where $g$ is the $SU(2)_L$ coupling constant and $\theta_W$ is the Weinberg angle. With the rotation matrix between field components and neutral eigenstates
defined in Eq.~\eqref{eq:matR}, we easily obtain ($i,j$ =1,2,3)
\be
|D_\mu \Phi_2|^2\,=\, \dots \,+\, \frac{g}{\cos\theta_W}\,\left(R_{ij} R_{ji} - R_{ii}
R_{jj}\right)\,Z_\mu\left(h_i\partial^\mu h_j \,-\,
h_j\partial^\mu h_i\right) \, .
\label{eq:zhihj}
\ee
Thus decays or production mechanisms of the form
$h_j \rightarrow Z\,h_i$, $Z\rightarrow h_j\,h_i$, for any $h_{i\neq j}$ dark neutral scalars, are {\em simultaneously}
possible (with the $Z$ boson possibly off-shell) which would clearly not be possible if the $h_i$ had definite CP
quantum numbers -- in fact, due to CP violation, the three dark scalars are neither CP-even nor CP-odd, but
rather states with mixed CP quantum numbers. The {\em simultaneous} existence of all $Z h_j\,h_i$ vertices, with
$i\neq j$, is a clear signal of CPV in the model, in clear opposition to what occurs, for instance,
in the CP-conserving 2HDM -- in that model $Z\rightarrow A\,h$ or $Z\rightarrow A\,H$ are
possible because $A$ is CP-odd and $h$, $H$ are CP-even, but $Z\rightarrow H\,h$ or $Z\rightarrow A\,A$ are
forbidden. Since in our model all vertices $Z h_j\,h_i$ with $i\neq j$ occur, the neutral scalars $h_i$
cannot have definite CP quantum numbers. Thereby CP violation is  established in the model in the dark sector.
Notice that no vertices of the form $Z h h_i$ are possible. This is not due to any CP
properties, however, but rather to the conservation of the $Z_2$ quantum number.
Thus observation of such decays or production mechanisms (all three possibilities for $Z\rightarrow h_j\,h_i$,
$i\neq j$, would have to be confirmed) could serve as confirmation of CPV in the model,
though the non-observability of the dark scalars would mean they would only contribute to missing energy
signatures.
Both at the LHC and at future colliders, hints on the existence of dark matter can appear in mono-$Z$
or mono-Higgs searches. The current model predicts cascade processes such as $q \bar{q} \, (e^+ e^-) \to Z^*
\to h_1 h_2 \to h_1 h_1 Z$ and  $q \bar {q} \, (e^+ e^-) \to Z^* \to h_1 h_2 \to h_1 h_1 h_{125}$, leading to
mono-Z and mono-Higgs events, respectively. This type of final states occurs in many dark matter models,
regardless of the CP-nature of the particles involved. Therefore, these are not good processes to probe
CP-violation in the dark sector.

\begin{figure}[t]
\centering
\includegraphics[height=4cm,angle=0]{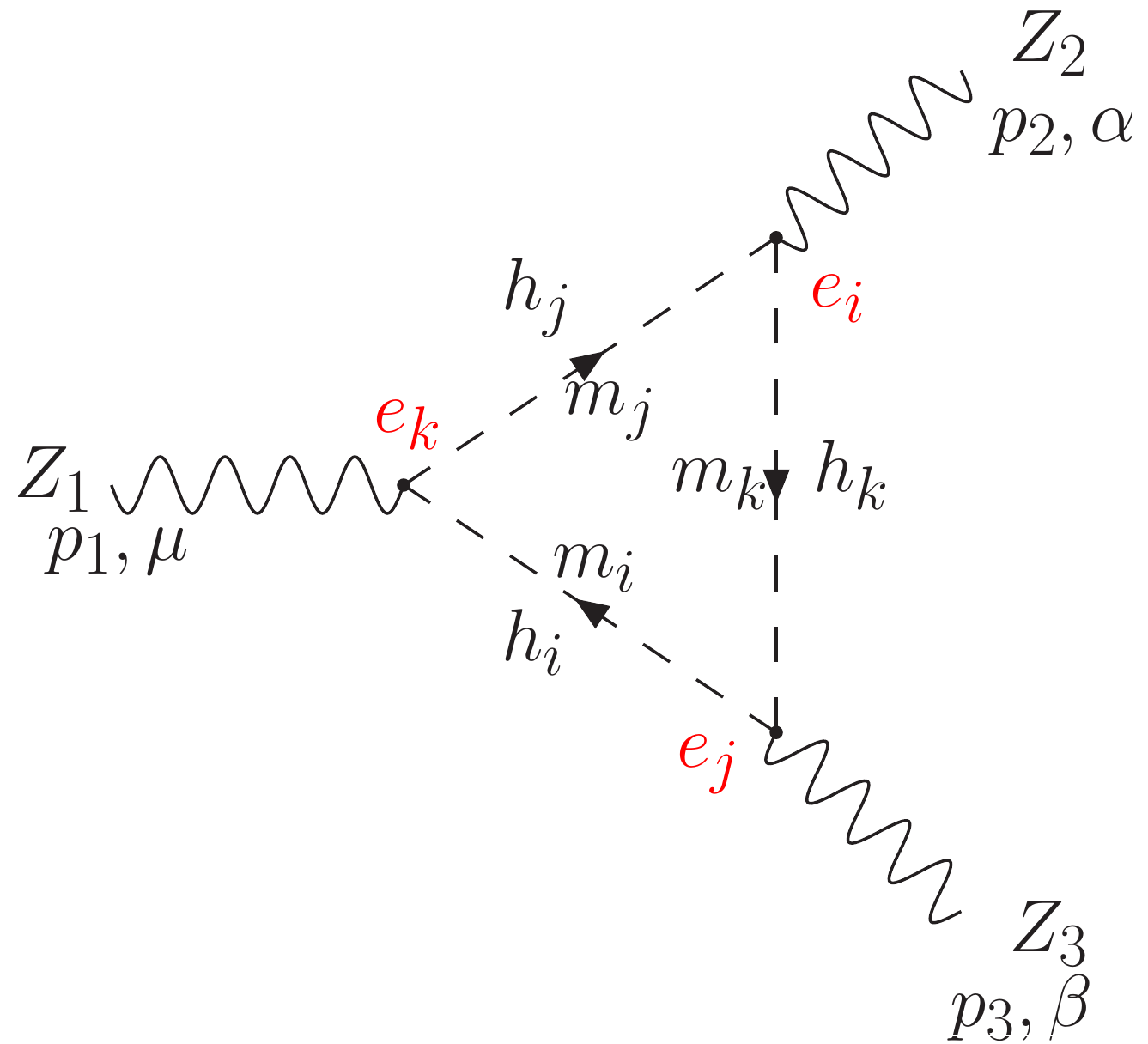}
\caption{Feynman diagram contributing to the CP violating form factor $f^Z_4$.
}
\label{fig:diag}
\end{figure}

However, though CPV occurs in the dark sector of the theory, it can have an observable impact on the
phenomenology of the SM particles. A sign of CPV in the model -- possibly the only type of signs of CPV which
might be observable
-- can be gleaned from the interesting work of Ref.~\cite{Grzadkowski:2016lpv} (see also
Ref.~\cite{Belusca-Maito:2017iob}), wherein 2HDM
contributions to the triple gauge boson vertices $ZZZ$ and $ZW^+W^-$ were considered. A Lorentz structure analysis
of the $ZZZ$ vertex, for instance~\cite{Hagiwara:1986vm,Gounaris:1999kf,Gounaris:2000dn,Baur:2000ae},
reveals that there are 14 distinct structures, which can be reduced to just two form factors on
the assumption of two on-shell $Z$ bosons and massless fermions, the off-shell $Z$ being produced by $e^+e^-$
collisions. Under these simplifying assumptions, the $ZZZ$ vertex function becomes ($e$ being the unit
electric charge)
\be
e\Gamma^{\alpha\beta\mu}_{ZZZ} \,=\,i\,e\,\frac{p_1^2 - m^2_Z}{m^2_Z} \left[f^Z_4 \left(p_1^\alpha g^{\mu\beta} +
p_1^\beta g^{\mu\alpha}\right)\,+\,f^Z_5 \epsilon^{\mu\alpha\beta\rho}
\left(p_2-p_3\right)_\rho\right] \;,
\label{eq:vert}
\ee
where $p_1$ is the 4-momentum of the off-shell $Z$ boson, $p_2$
and $p_3$ those of the remaining (on-shell) $Z$ bosons.
The dimensionless $f_4^Z$ form factor is CP violating, but the $f_5^Z$
coefficient preserves CP.
In our model there is only one-loop diagram contributing to this form factor,
shown in Fig.~\ref{fig:diag}. As can be inferred from the diagram
there are three different neutral scalars
circulating in the loop -- in fact, the authors of Ref.~\cite{Grzadkowski:2016lpv} showed that in the 2HDM with
explicit CPV (the C2HDM) the existence of at least three neutral scalars with different CP quantum numbers
that mix among themselves is a necessary condition for non-zero values for $f^Z_4$.
Notice that in the C2HDM there are {\em three} diagrams contributing to $f^Z_4$ -- other than the diagram
shown in Fig.~\ref{fig:diag}, the C2HDM calculation involves an additional diagram with an internal $Z$ boson
line in the loop, and another, with a neutral Goldstone boson $G^0$ line in the loop. In
our model, however, the discrete $Z_2$ symmetry we imposed forbids the
vertices $Z Z h_j$ and $Z G^0 h_i$ (these vertices do occur in the C2HDM, being allowed by that model's
symmetries), and therefore those two additional diagrams are identically zero.
In~\cite{Grzadkowski:2016lpv} an expression for $f^Z_4$ in the C2HDM was found, which can easily be adapted to our model, by only keeping the contributions corresponding to the diagram of Fig.~\ref{fig:diag}. This results
in
\be
f^Z_4(p_1^2) \,=\, -\,\frac{2\alpha}{\pi s^3_{2\theta_W}}\,\frac{m^2_Z}{p_1^2 - m^2_Z}\,f_{123}\;
\sum_{i,j,k}\,\epsilon_{ijk}\,C_{001}(p_1^2,m^2_Z,m^2_Z,m^2_i,m^2_j,m^2_k) \, ,
 \label{eq:f4Z}
\ee
where $\alpha$ is the electromagnetic coupling constant and the
{\tt LoopTools}~\cite{Hahn:1998yk} function $C_{001}$
is used. The $f_{123}$ factor denotes the product of the couplings from three different vertices,
given in Ref.~\cite{Grzadkowski:2016lpv} by
\be
f_{123}\,=\,\frac{e_1 e_2 e_3}{v^3}\,,
\ee
where the $e_{i,j,k}$ ($i,j,k=1,2,3$) factors, shown in
Fig.~\ref{fig:diag}, are related to the coupling coefficients that appear
in the vertices $Z h_i h_j$ (in the C2HDM they also concern the $ZG^0 h_i$ and $ZZ h_i$
vertices, {\it  cf.}~\cite{Belusca-Maito:2017iob}).  With the
conventions of the current paper, we can extract
these couplings from Eq.~\eqref{eq:zhihj} and it is easy to show that
\ba
f_{123} & = & \left(R_{12} R_{21} - R_{11}R_{22}\right)\, \left(R_{13} R_{31} - R_{11}
R_{33}\right) \, \left(R_{23} R_{32} - R_{22} R_{33}\right) \nonumber \\
& = & R_{13} R_{23} R_{33}\,,
\label{eq:f123}
\ea
where the simplification that led to the last line originates from the orthogonality of the $R$ matrix.
We observe that the maximum value that $f_{123}$ can assume is $(1/\sqrt{3})^3$, corresponding
to the {\em maximum mixing} of the three neutral components, $\rho$,
$\eta$ and $\Phi_S \equiv s$.
This is quite different from what one expects to happen in the C2HDM, for instance -- there one
of the mixed neutral states is the observed 125 GeV scalar, and its properties are necessarily
very SM-like, which implies that the $3\times 3$ matrix $R$ should approximately have the form
of one diagonal element with value close to 1, the corresponding row and column with elements very small
and a $2\times 2$ matrix mixing the other eigenstates\footnote{Meaning, a neutral scalar mixing very similar
to the CP-conserving 2HDM, where $h$ and $H$ mix via a $2\times 2$ matrix but $A$ does not mix with the
CP-even states.}. Within our model, however, the three neutral
dark fields can mix as much or as little as possible.

\begin{figure}[t]
\centering
\includegraphics[height=8cm,angle=0]{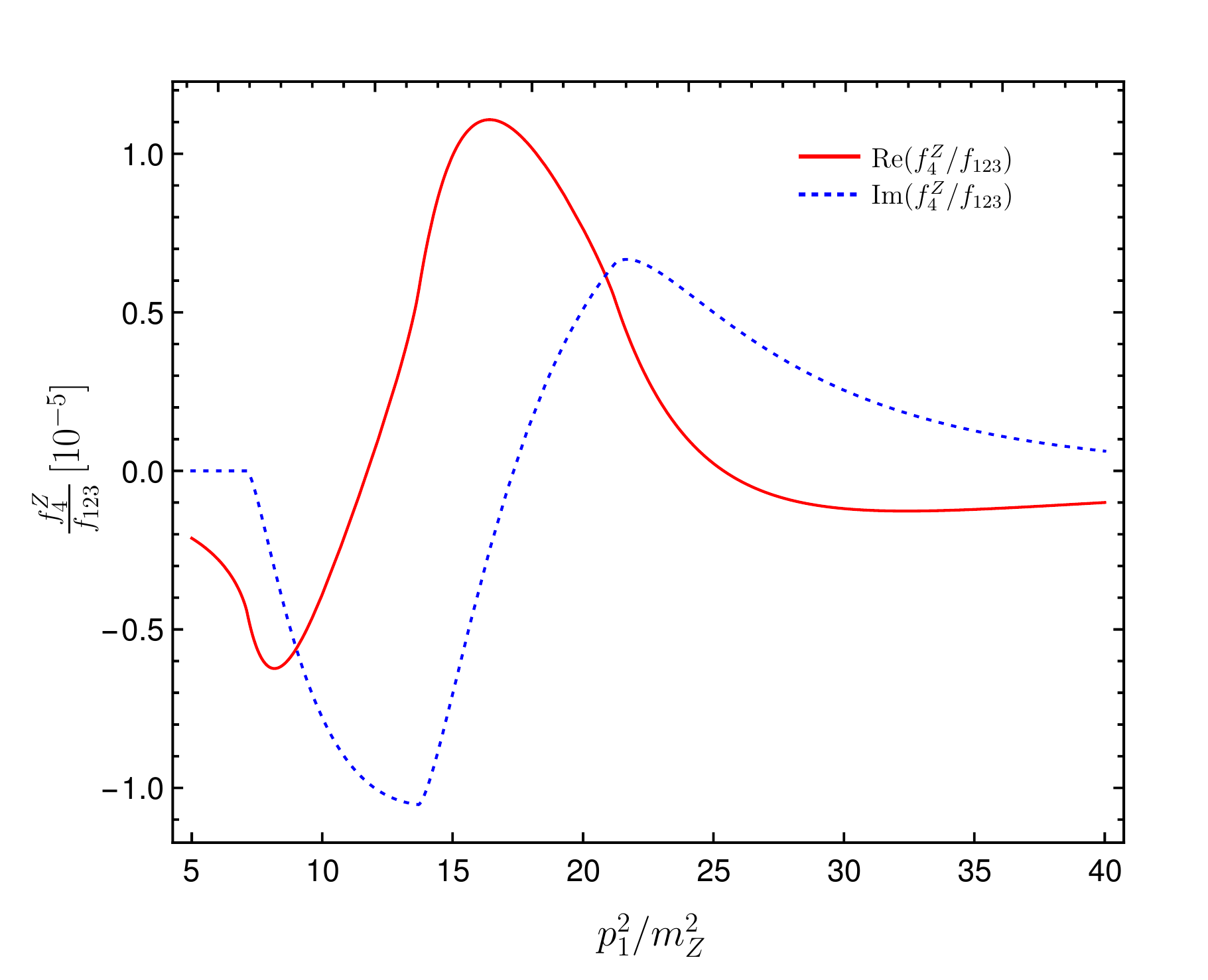}
\caption{The CP-violating $f_4^Z(p_1^2)$ form factor, normalized
    to $f_{123}$, for $m_{h_1}= 80.5$ GeV, $m_{h_2}=162.9$ GeV and
$m_{h_3}=256.9$ GeV, as a function of the squared off-shell $Z$ boson
4-momentum $p_1^2$, normalized to $m_Z^2$.
}
\label{fig:f4}
\end{figure}

\begin{figure}[t]
\begin{tabular}{cc}
\includegraphics[height=7cm,angle=0]{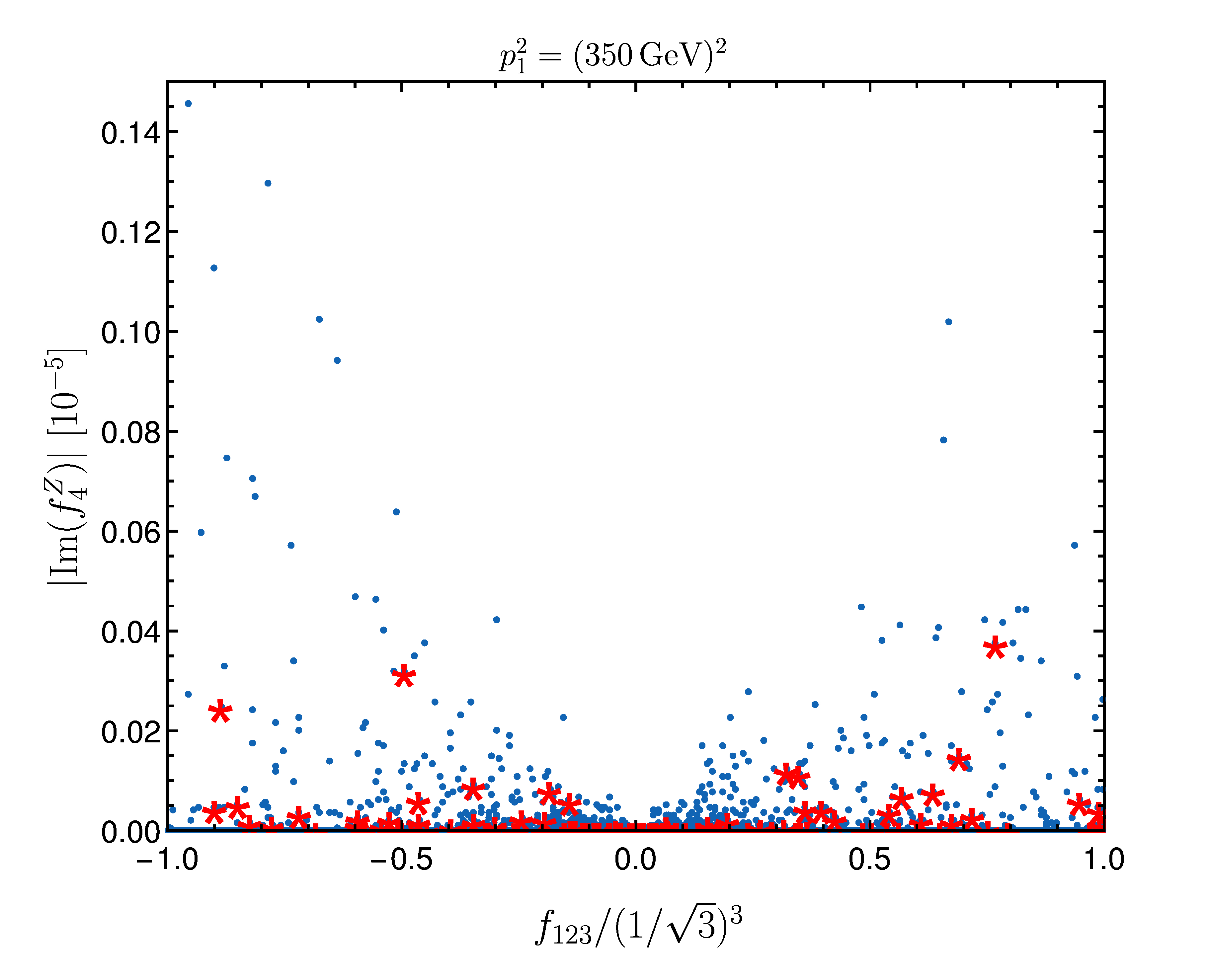}&
\includegraphics[height=7cm,angle=0]{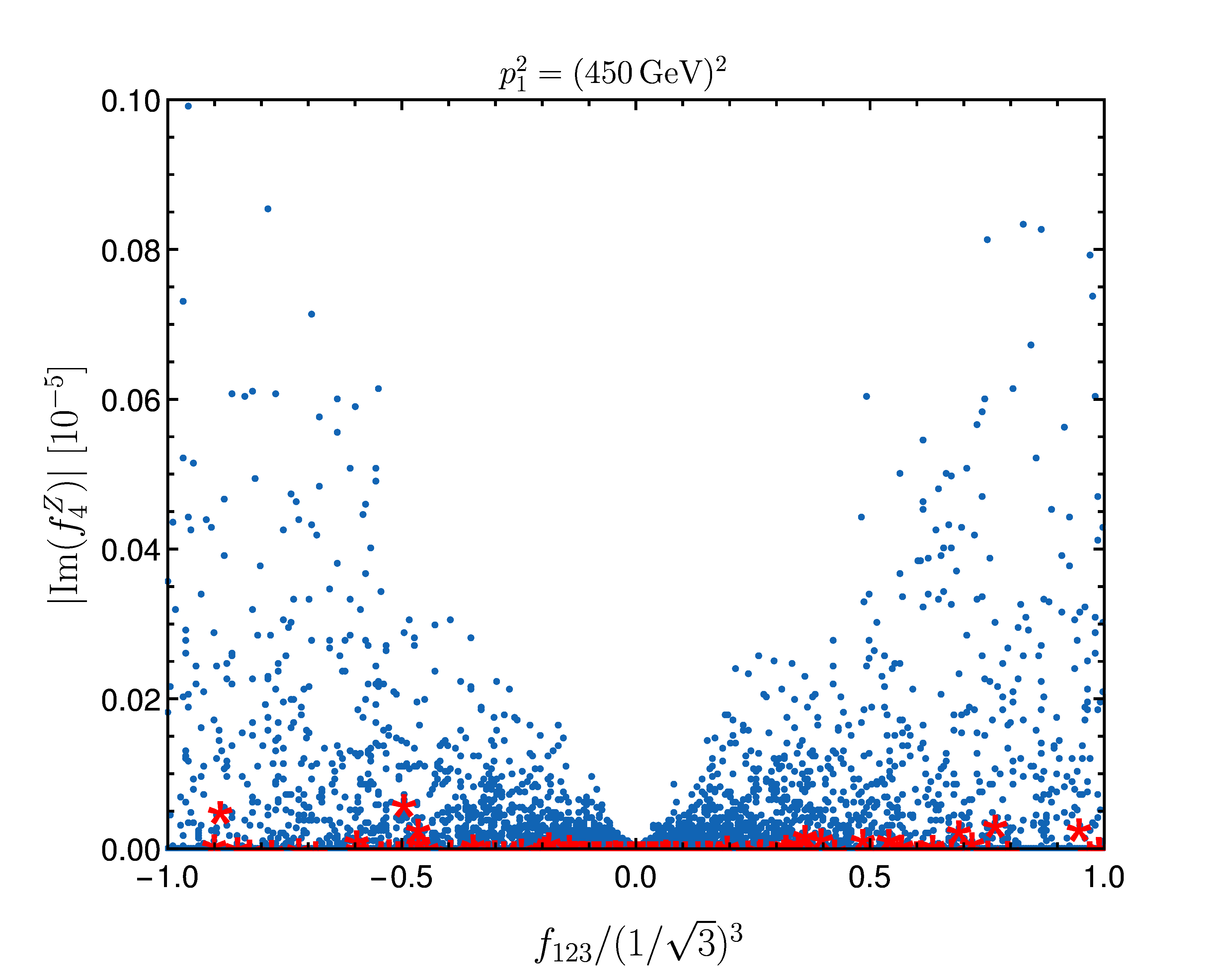}\\
 (a) & (b)
\end{tabular}
\caption{Scatter plots for the imaginary part of $f_4^Z$ as a function of the combined
$Z$-scalars coupling $f_{123}$ of Eq.~\eqref{eq:f123}, divided by its maximum possible value of $(1/\sqrt{3})^3$.
In (a) results for $p_1^2$ = $($350 GeV$)^2$; in (b),
$p_1^2$ = $($450 GeV$)^2$.
In red, points for which the masses of all the dark scalars are
smaller than 200~GeV, $m_{h_i}<200$ GeV ($i = 1,2,3$).
}
\label{fig:imf4}
\end{figure}

In Fig.~\ref{fig:f4} we show, for a random combination of dark scalar masses ($m_{h_1} \simeq 80.5$ GeV,
$m_{h_2} \simeq 162.9$ GeV and $m_{h_3} \simeq 256.9$ GeV) the evolution
of $f_4^Z$ normalized to $f_{123}$~\footnote{For this specific parameter space point, we have
$f_{123} \simeq -0.1835$.} with $p_1^2$, the
4-momentum of the off-shell $Z$ boson. This can be compared with
Fig.~2 of Ref.~\cite{Grzadkowski:2016lpv}, where
we see similar (if a bit larger) magnitudes for the real and imaginary parts of $f_4^Z$,
despite the differences in masses for the three neutral scalars in both situations (in that figure,
the masses taken for $h_1$ and
$h_3$ were, respectively, 125 and 400 GeV, and several values for the $h_2$ mass
were considered). As can be
  inferred from Fig.~\ref{fig:f4}, $f_4^Z$ is at most of the order of $\sim 10^{-5}$.
  For the parameter scan described in the previous section, we obtain, for the
  imaginary part of $f_4^Z$, the values shown in Fig.~\ref{fig:imf4}. We considered
two values of $p_1^2$ (corresponding to two possible collision energies for a future linear
collider). The imaginary part of $f_4^Z$ (which, as we will see, contributes directly to
CP-violating observables such as asymmetries) is presented as a function of the overall coupling $f_{123}$ defined
in Eq.~\eqref{eq:f123}. We in fact present results as a function of $f_{123}/(1/\sqrt{3})^3$, to
illustrate that indeed the model perfectly allows maximum mixing between the neutral, dark scalars.
Fig.~\ref{fig:imf4} shows that the maximum values for $|$Im$(f^Z_4)|$ are reached
for the maximum mixing scenarios. We also highlight in red the points for which the dark
neutral scalars $h_i$ have masses smaller than 200 GeV.
The loop functions in the definition of $f_4^Z$, Eq.~\eqref{eq:f4Z}, have a complicated
dependence on masses (and external momentum $p_1$) so that an analytical demonstration is not
possible, but the plots of Fig.~\ref{fig:imf4} strongly imply that choosing all dark scalar
masses small yields smaller values for $|$Im$(f^Z_4)|$. Larger masses, and larger
mass splittings, seem to be required for larger $|$Im$(f^Z_4)|$.
A reduction on the maximum values of $|$Im$(f^Z_4)|$ (and
$|$Re$(f^Z_4)|$) with increasing external momentum is observed (though
that variation is not linear, as can be appreciated from
Fig.~\ref{fig:f4}). A reduction of the maximum values of $|$Im$(f^Z_4)|$ (and
$|$Re$(f^Z_4)|$) when the external momentum
tends to infinity is also observed.

The smaller values for $|$Im$(f^Z_4)|$ for the red points
can be understood in analogy with the 2HDM. The authors of Ref.~\cite{Grzadkowski:2016lpv}
argue that the occurrence of CPV in the model implies a non-zero value for the basis-invariant
quantities introduced in Refs.~\cite{Lavoura:1994fv,Botella:1994cs}, in particular for the
imaginary part of the $J_2$ quantity introduced therein. Since Im$(J_2)$ is proportional to the product
of the differences in mass squared of all neutral scalars, having all those scalars with lower masses
and lower mass splittings reduces Im$(J_2)$ and therefore the amount of CPV in the model. Now, in
our model the CPV basis invariants will certainly be different from those of the 2HDM, but we can adapt
the argument to understand the behaviour of the red points in Fig.~\ref{fig:imf4}: those red points
correspond to three dark neutral scalars with masses lower than 200 GeV, and therefore their mass splittings
will be small (compared to the remaining parameter space of the model). In the limiting case of three
degenerate dark scalars, the mass matrix of Eq.~\eqref{eq:mn} would be proportional to the identity
matrix and therefore no mixing between different CP states would occur. With this analogy, we can understand
how regions of parameter space with larger mass splittings between the dark neutral scalars
tend to produce larger values of $|$Im$(f^Z_4)|$.

Experimental collaborations have been probing double-$Z$ production to look for anomalous
couplings such as those responsible for a $ZZZ$ vertex~\cite{Aaltonen:2008mv,Aaltonen:2009fd,
Abazov:2011td,Abazov:2012cj,Aad:2011xj,Chatrchyan:2012sga,Aad:2012awa,CMS:2014xja,Khachatryan:2015pba}.
The search for anomalous couplings in those works uses the effective
Lagrangian for triple neutral
vertices proposed in Ref.~\cite{Hagiwara:1986vm}, parametrised as
\be
\mathcal{L}_{\mathrm{V}ZZ}  = -\frac{e}{m_Z^2} \left\{
\left[f_4^\gamma\left(\partial_\mu F^{\mu\alpha}\right)
+f_4^Z\left(\partial_\mu Z^{\mu\alpha}\right)\right]
Z_\beta\left(\partial^\beta Z_\alpha\right)
-\left[f_5^\gamma\left(\partial^\mu F_{\mu\alpha}\right)
+f_5^Z\left(\partial^\mu Z_{\mu\alpha}\right)\right]
\tilde{Z}^{\alpha\beta}Z_\beta
\right\},
\label{eq:effZZZ}
\ee
where $\gamma ZZ$ vertices were also considered.
In this equation, $F_{\mu\nu}$ is the electromagnetic tensor,
$Z_{\mu\nu} = \partial_\mu Z_\nu - \partial_\nu Z_\mu$ and
$\tilde{Z}_{\mu\nu} = \epsilon_{\mu\nu\rho\sigma} Z^{\rho\sigma}/2$.
The $f_4^Z$ coupling above is taken to be a constant,
and as such it represents at most an approximation to the
$f_4^Z(p_1^2)$ of Eq.~\eqref{eq:f4Z}. Further,
the analyses of the experimental collaborations mentioned above take this coupling to be real, whereas
the imaginary part of $f_4^Z(p_1^2)$ is the quantity of interest in many interesting observables.
With all that under consideration, latest results from LHC~\cite{Khachatryan:2015pba}
already probe the $f_4^Z$ coupling of Eq.~\eqref{eq:effZZZ} to order $\sim 10^{-3}$, whereas the typical
magnitude of $f_4^Z(p_1^2)$ (both real and imaginary parts) is $\sim
10^{-5}$. We stress , however, that
the two quantities cannot be directly compared, as they represent very different approaches to the
$ZZZ$ vertex. A thorough study of the experimental results of~\cite{Khachatryan:2015pba} using the
full expression for the $ZZZ$ vertex of Eq.~\eqref{eq:vert} and the full momentum (and scalar masses)
dependence of the form factors is clearly necessary, but beyond the scope of the current work.

\begin{figure}[t]
\centering
\includegraphics[height=7cm,angle=0]{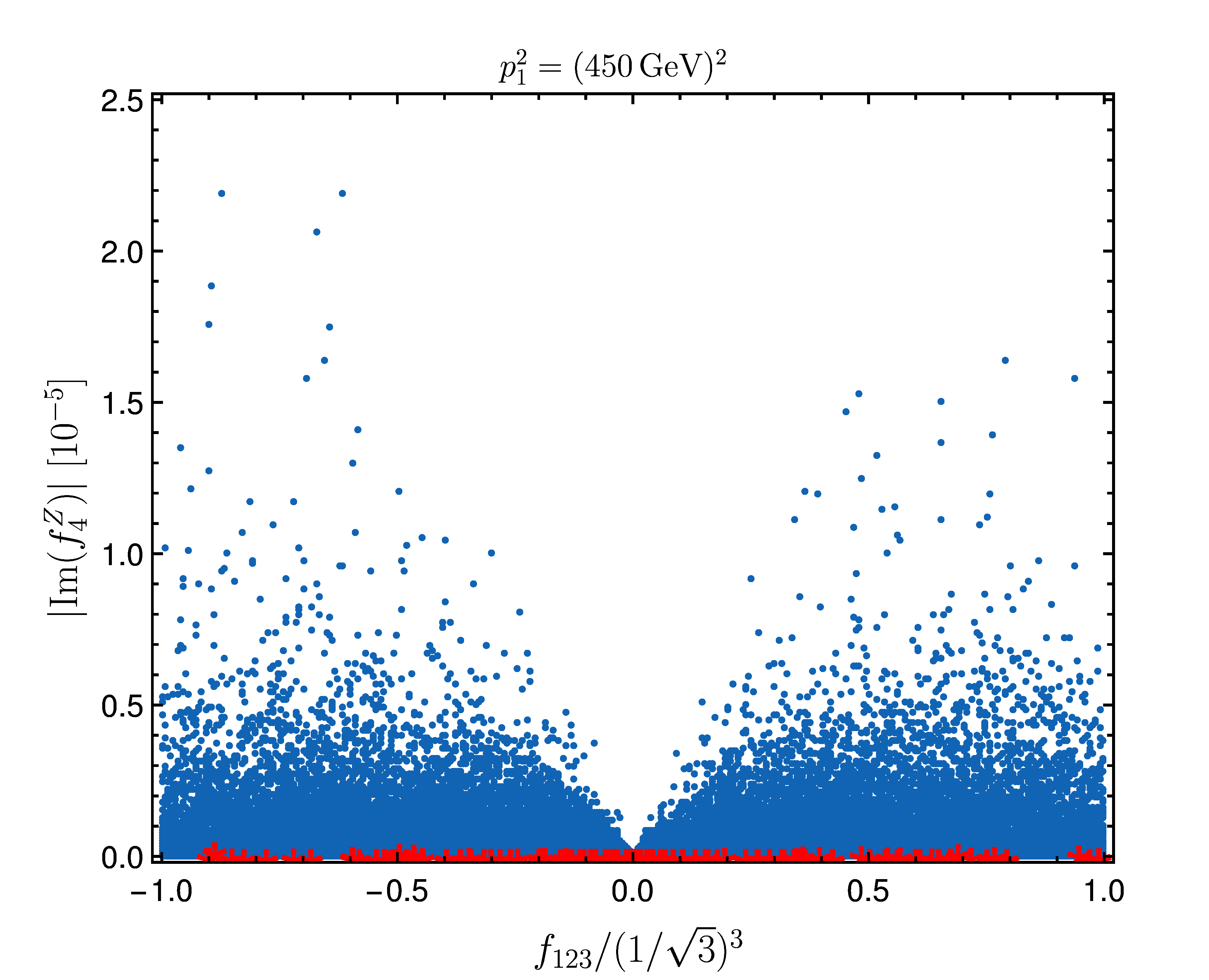}
\caption{Scatter plot for the imaginary part of $f_4^Z$ for the $ZW^+W^-$ vertex from
Eq.~\eqref{eq:f4ZW}, as a function of the combined
$Z$-scalars coupling $f_{123}$, divided by its maximum possible value of $(1/\sqrt{3})^3$.
The external $Z$ boson 4-momentum is
$p_1^2$ = $($450 GeV$)^2$.
In red, points for which the masses of all the dark neutral scalars are
smaller than 200~GeV, $m_{h_i}<200$ GeV ($i = 1,2,3$).
}
\label{fig:imf4W}
\end{figure}

The crucial aspect to address here, and the point we wish to make with the present section, is that
$f_4^Z(p_1^2)$ is non-zero in the model under study in this paper.
Despite the fact that the neutral scalars
contributing to the form factor are all dark particles, CP
violation is therefore present in the model and it can indeed be ``visible" to us, having consequences
in the non-dark sector.
We also analysed other vertices, such as $ZW^+W^-$ -- there CPV form factors also arise, also
identified as ``$f_4^Z$", and for our
parameter scan we computed it by once again adapting the results of Ref.~\cite{Grzadkowski:2016lpv}
to our model. In the C2HDM three Feynman diagrams contribute to this CP-violating form factor (see Fig. 17
in~\cite{Grzadkowski:2016lpv}) but in our model the $Z_2$ symmetry eliminates the vertices $h_i W^+ W^-$
and $h_i G^+ W^-$, so only one diagram involving the charged scalar survives. From Eq. (4.4) of
Ref.~\cite{Grzadkowski:2016lpv}, we can read the expression of the CP-violating form factor $f_4^Z$
from the $ZW^+W^-$ vertex, obtaining
\be
f^Z_4(p_1^2) \,=\, \frac{\alpha}{\pi s^2_{2\theta_W}}\,f_{123}\;
\sum_{i,j,k}\,\epsilon_{ijk}\,C_{001}(p_1^2,m^2_W,m^2_W,m^2_i,m^2_j,m^2_{H^+}) \, .
 \label{eq:f4ZW}
\ee
Interestingly, this form factor is larger, by roughly a factor of ten, than the corresponding
quantity in the $ZZZ$ vertex (though still smaller than the corresponding C2HDM typical values).
This is illustrated in Fig.~\ref{fig:imf4W}, where we plot
the imaginary part of $f^Z_4$ as given by Eq.~\eqref{eq:f4ZW} for $p_1^2 = ($450 GeV$)^2$,
having obtained non-zero values. Therefore CPV also occurs in the $ZW^+W^-$ interactions in this
model, though presumably it would be no easier to experimentally
establish than for the $ZZZ$ vertex. The point we wished to make does not change, however --
if even a single non-zero CPV quantity is found, then CP violation occurs in the model.

\begin{figure}[t]
\centering
\includegraphics[height=8cm,angle=0]{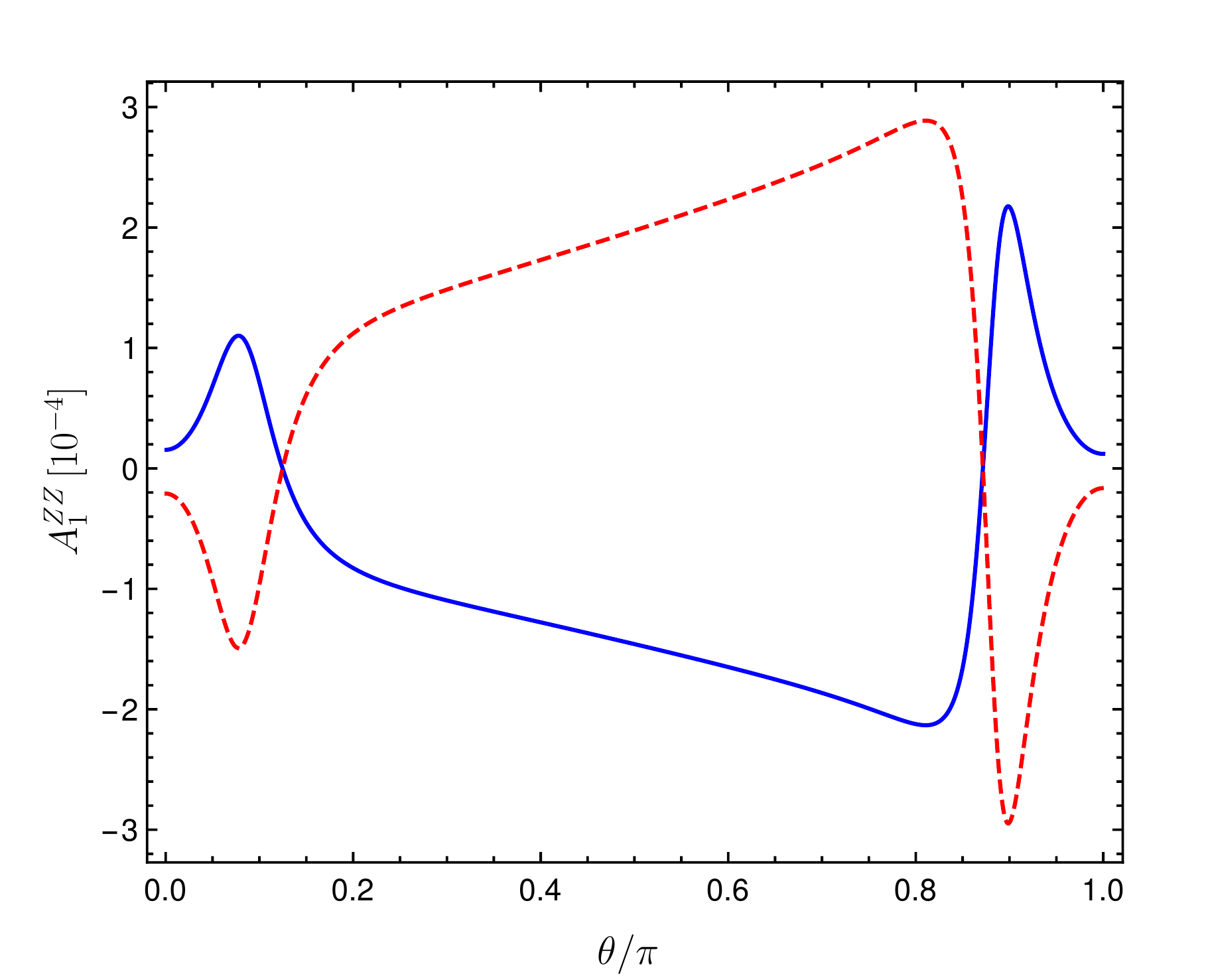}
\caption{The $A_1^{ZZ}$ asymmetry of Eq.~\eqref{eq:A1ZZ} as a function of the angle $\theta$. The
blue (full) curve corresponds to the largest positive value of
$\mbox{Im}\left(f_4^Z(p_1^2)\right)$ in
our parameter scan, the red (dashed) one to the smallest negative value for the same quantity. In
both cases, $p_1^2 = (450$ GeV$)^2$.
}
\label{fig:asym}
\end{figure}

As an example of a possible experimental observable to which the form factors $f_4^Z$ for the
$ZZZ$ interactions might contribute,
let us take one of the asymmetries considered in Ref.~\cite{Grzadkowski:2016lpv}, using the techniques of
Ref.~\cite{Chang:1994cs}.
Considering a future linear collider and the process $e^+e^-\rightarrow ZZ$, taking
cross sections for unpolarized beams $\sigma_{\lambda,\bar{\lambda}}$
for the production of two $Z$
bosons of helicities $\lambda$ and $\bar{\lambda}$ (assuming the
helicity of the $Z$ bosons can be determined),
the asymmetry $A_1^{ZZ}$ is defined as
\be
A_1^{ZZ}\,=\,\frac{\sigma_{+,0} - \sigma_{0,-}}{\sigma_{+,0} + \sigma_{0,-}}\,=\,
-4\beta \gamma^4 \left[(1 + \beta^2)^2 - 4 \beta^2\cos^2\theta\right]\,
{\cal F}_1(\beta,\theta)\,\mbox{Im} \left(f_4^Z(p_1^2)\right)\,,
\label{eq:A1ZZ}
\ee
with $\theta$ the angle between the electron beam and the closest $Z$
boson with positive helicity, $\beta = \sqrt{1 - 4 m^2_Z/p_1^2}$ denoting the velocity
of the produced $Z$ bosons and the function ${\cal F}_1(\beta,\theta)$ is given in appendix D of
Ref.~\cite{Grzadkowski:2016lpv}. Choosing the
two points in our parameter scan with largest (positive) and smallest (negative) values of
$\mbox{Im} \left(f_4^Z(p_1^2)\right)$ for $p_1^2 = (450$ GeV$)^2$, we
obtain the two curves shown in Fig.~\ref{fig:asym}.
Clearly, the smallness ($\sim 10^{-5}$) of the $f_4^Z$ form factor renders the value of this asymmetry
quite small, which makes its measurement challenging.
This raises the possibility that asymmetries involving the $ZW^+W^-$ vertex might be easier
to measure than those pertaining to the $ZZZ$ anomalous interactions, since we have shown that
$f^Z_4$ is typically larger by a factor of ten in the former vertex compared to the latter one.
To investigate this possibility, we compared $A_1^{ZZ}$, considered above, with the $A_1^{WW}$ asymmetry
defined in Eq. (5.21) of Ref.~\cite{Grzadkowski:2016lpv}.
A direct comparison of the maximum values of $A_1^{WW}$ and $A_1^{ZZ}$
shows that for some regions of parameter space the former quantity can indeed be one order
of magnitude larger than the latter one; but that is by no means a generic feature, since for other
choices of model parameters both asymmetries can also be of the same order. Notice that both asymmetries show a quite different $\sqrt{s}$ dependence.

\section{Conclusions}
\label{sec:conc}

We presented a model whose scalar sector includes two Higgs doublets and a real singlet. A specific region
of parameter space of the model yields a vacuum which preserves a discrete symmetry imposed on
the model -- thus a charged scalar and three neutral ones have a ``dark" quantum number preserved in
all interactions and have no interactions with fermions. The lightest of them, chosen
to be a neutral particle,
is therefore stable and a good dark matter candidate. The first doublet yields the necessary Goldstone
bosons and a neutral scalar which has automatically a behaviour almost
indistinguishable from the SM Higgs boson.
A parameter scan of the model, imposing all necessary theoretical and experimental constraints (including
bounds due to relic density and dark matter searches, both direct and indirect) shows that the
SM-like scalar state indeed complies with all known LHC data for the Higgs boson -- some deviations may
occur in the diphoton signal rate due to the extra contribution of a charged scalar to the involved
decay width, but we have shown such deviations are at most roughly 20\% of the expected SM result when all
other constraints are satisfied, and this is still well within current
experimental uncertainties.

The interesting thing about the model presented in this paper is the occurrence of explicit CP violation
exclusively within the dark matter sector. A complex phase allowed in the potential forces the neutral
components of the second (dark) doublet to mix with the real singlet to yield three neutral eigenstates,
none of which possesses definite quantum numbers. Signals of this CP violation would not be observed in
the fermion sector (which, by the way, we assume is identical to the
one of the SM, and therefore has the usual
CKM-type source of CP violation) nor in the interactions of the SM-like scalar -- protected as it is
by the unbroken $Z_2$ symmetry, and by the mass ranges chosen for the dark scalars, $h$ will
behave like a purely CP-even SM-like scalar, even though the CP symmetry of the model is explicitly broken
in the scalar sector as well! Can the model then be said to be CP violating at all? The answer is yes,
as an analysis of the contributions from the dark sector to the $ZZZ$ vertex demonstrates. Even though the
dark particles have no direct fermion interactions and could elude detection, their presence could
be felt through the emergence of anomalous triple gauge boson vertices. Though we concentrated
mainly on $ZZZ$ vertices we also studied $ZW^+W^-$ interactions, but our main purpose was to
show CPV is indeed occurring. Direct measurements of experimental observables probing this CPV are
challenging: we have considered a specific asymmetry, $A_1^{ZZ}$, built with $ZZ$ production cross
sections, but
the magnitude of the CPV form factor $f_4^Z$ yields extremely small values for that asymmetry, or indeed
for other such variables we might construct. Direct measurements of $ZZ$ production cross sections
could in theory be used to constraint anomalous $ZZZ$ vertex form factors -- and indeed several
experimental collaborations, from LEP, Tevatron and LHC, have tried that. But the experimentalists'
approach is based on constant and real form factors, whereas model-specific expressions for
$f_4^Z$ such as those considered in our work yield quantities highly dependent on external momenta,
which boast sizeable imaginary parts as well. Thus a direct comparison with current experimental
analyses is not conclusive.

The other remarkable fact is the amount of ``damage" the mere inclusion of a real singlet can do to
the model with two doublets. As repeatedly emphasised in the text, the model we considered is very similar
to the Inert 2HDM -- it is indeed simply the IDM with an added real singlet and a tweaked discrete symmetry,
extended to the singlet having a ``dark charge" as well. But whereas CP violation --
explicit or spontaneous --
is entirely impossible within the scalar sector of the IDM, the presence of the extra singlet produces
a completely different situation. That one obtains a model with explicit CPV is all the more remarkable
when one considers
that the field we are adding to the IDM is a {\em real} singlet, not even a complex one. Notice that within
the IDM it is even impossible to tell which of the dark neutral scalars is CP-even and which is CP-odd -- all
that can be said is that those two eigenstates have opposite CP quantum numbers. The addition of a real
singlet completely changes the CP picture.

The occurrence of CP violation in the dark matter sector can be simply
a matter of curiosity, but one should not
underestimate the possibility that something novel might arise from it. If the current picture
of matter to dark matter
abundance is indeed true and the observed matter is only 5\% of the total content of the universe,
then one can speculate how CP violation occurring in the interactions of the remainder
matter might have affected
the cosmological evolution of the universe. We reserve such studies for a follow-up work.

\section*{Acknowledgements}
We  acknowledge  the  contribution  of  the research  training  group  GRK1694 `Elementary  particle  physics
at  highest energy  and  highest  precision’.
PF and RS are supported in part by the National Science Centre, Poland, the
HARMONIA project under contract UMO-2015/18/M/ST2/00518.
JW gratefully acknowledges funding from the PIER Helmholtz Graduate School.


\bibliography{CPDark}

\end{document}